\input{aipcheck}

\documentclass[
    ,final            
  ]
  {aipproc}

\layoutstyle{8x11single}

\begin{document}

\title{A realistic renormalizable supersymmetric $E_6$ model\footnote{Talk given by Borut Bajc.}}


\author{Borut Bajc}{
  address={J. Stefan Institute, 1000 Ljubljana, Slovenia}
  ,altaddress={Faculty of Mathematics and Physics, University of Ljubljana, 1000 Ljubljana, Slovenia} 
}

\author{Vasja Susi\v{c}}{
  address={J. Stefan Institute, 1000 Ljubljana, Slovenia}
}

\begin{abstract}
A complete realistic model based on the supersymmetric version of $E_6$ is presented. It consists of
three copies of matter $27$, and a Higgs sector made of $2\times(27+\overline{27})+351'+\overline{351'}$
representations. An analytic solution to the equations of motion is found which spontaneously breaks
the gauge group into the Standard Model. The light fermion mass matrices are written down explicitly
as non-linear functions of three Yukawa matrices. This contribution is based on Ref. \cite{mi}.
\end{abstract}

\maketitle

\newcommand{\beq}{\begin{equation}}
\newcommand{\eeq}{\end{equation}}
\newcommand{\bea}{\begin{eqnarray}}
\newcommand{\eea}{\end{eqnarray}}
\newcommand{\bem}{\begin{pmatrix}}
\newcommand{\eem}{\end{pmatrix}}
\newcommand{\noi}{\noindent}
\def\SO{\mathrm{SO}}
\def\SU{\mathrm{SU}}
\def\RED{{}}
\def\ydt{\widetilde{Y}_{27}}
\def\ytt{\tilde Y_{\overline{351}'}}
\def\yd{Y_{27}}
\def\yt{Y_{\overline{351}'}}

\section{Introduction}

Among the different relevant grand unifying theories (GUT) \cite{Georgi:1974sy} the one based on the $E_6$ group \cite{Gursey:1975ki} is probably
the least studied. In particular, a complete treatment of the Higgs sector with explicit vacuum solutions is still missing.
We fill this gap by presenting the first complete realistic model based on the supersymmetric version of $E_6$. This
choice is motivated by the theoretical successes of supersymmetry on one side and by simplicity on the other (the
Higgs sector is simpler in supersymmetry).

Let's first try to see which representations are necessary to break $E_6$ into the Standard Model (SM) gauge group
SU(3)$\times$SU(2$\times$U(1). Among the irreducible representations of $E_6$ the smallest one is the $27$-dimensional.
This seems the ideal candidate for matter, however having only SU(5) singlets obviously cannot serve the purpose of
breaking the gauge group all the way to the SM. A special place is then reserved for $351'$. In fact it contains 5 singlets
under the Standard Model gauge group, as well as the two crucial SO(10) Higgses
in the 10 and 126 dimensional representations \cite{Babu:1992ia,Aulakh:2003kg}. It is thus a good starting point to consider a theory with the $351'+\overline{351}'$
Higgses. We will see that these two fields can break the most general renormalizable $E_6$-symmetric superpotential
only to the Pati-Salam subgroup SU(4)$_C\times$SU(2)$_L\times$SU(2)$_R$. It will be shown explicitly that the choice of
$351'+\overline{351}'+27+\overline{27}$ is just enough for the goal: the vacuum solution keeps 12 gauge bosons massless.

The next issue for a realistic model is to identify the minimal supersymmetric standard model (MSSM) Higgses. In doing that one should find a light weak doublet pair and
at the same time get rid of all color triplets. This is the infamous doublet-triplet splitting problem, unfortunately present in most
grand unified theories, which is solved in all known minimal models by a simple fine-tuning of the potential parameters. Although
the previous Higgs choice includes a large number of fields with the quantum numbers of the MSSM Higgs doublets, we will show that no fine-tuning among the
parameters is possible without having at the same time also massless color triplet states, which would mediate phenomenologically
unacceptable fast proton decay. A plausible reason for that can be identified with the inability for the same parameters to
describe both symmetry breaking and doublet-triplet splitting fine-tuning. This is evaded by introducing an extra
$\widetilde{27}+\overline{\widetilde{27}}$: the new parameters allow for the doublet-triplet fine-tuning, so that the MSSM Higgses
live in these tilde fields.

The last step is to study the form of the Yukawa matrices. Since the MSSM Higgses do not live in $\overline{351}'$ and $27$, it
seems at first sight that only one Yukawa matrix is available, which would imply no mixing as well as the equality of the down quark and
charged lepton masses. A more careful investigation however shows the presence
of vector-like matter in the three generations of matter $27_F$ of $E_6$. In practice the matter $\bar 5$ of SU(5) can live both in 16 and 10
of SO(10), so the orthogonal component is heavy: once these three extra $\bar 5$'s are integrated out, they correct the
naive one-Yukawa picture we mentioned before. We work out in detail the form of the Yukawa matrices. A parameter counting shows
that there are enough parameters to describe all the masses and mixings of the Standard Model, neutrino included.

\section{Elements of $E_6$ group theory}

Similarly to the $\SU(N)$ groups, the $E_6$ group has two type of tensor indices: the upper or fundamental indices and the lower or antifundamental indices.
Both types of indices go from $1$ to $27$, which is the dimensionality of the fundamental and antifundamental representations.
Higher dimensional irreducible representations can then be constructed as tensors with these indices, satisfying extra
constraints like simmetricity or antisymmetricity. Finally, similarly to the case of the completely antisymmetric $\SU(N)$ invariant Levi-Civita tensor $\epsilon_{\alpha_1\ldots\alpha_N}$ or $\epsilon^{\alpha_1\ldots\alpha_N}$, we have in $E_6$ a 3-index completely symmetric
invariant tensor $d_{\mu\nu\lambda}$ (and the numerically equivalent $d^{\mu\nu\lambda}$) with $\mu,\nu,\lambda=1\ldots27$.

The representations used in our model are the following:
\begin{eqnarray}
27^\mu&\ldots&{\rm fundamental,}\\
\overline{27}_\mu&\ldots&{\rm anti-fundamental,}\\
351'^{\mu\nu}=+351'^{\nu\mu}&\ldots&{\rm two\;indices\;symmetric}\; (d_{\lambda\mu\nu}351'^{\mu\nu}=0),\\
\overline{351'}_{\mu\nu}=+\overline{351'}_{\nu\mu}&\ldots&{\rm two\;indices\;symmetric}\; (d^{\lambda\mu\nu}\overline{351'}_{\mu\nu}=0).
\end{eqnarray}
Using the normalization
\begin{eqnarray}
d_{\mu\lambda\rho} d^{\lambda\rho\nu}&=&10 \delta_{\mu}{}^{\nu},
\end{eqnarray}
the $d$-tensor contains only values $0$, $-1$ or $1$. Another important property of this tensor is the fact that it gives zero, as soon as two of its indices take the same value. This property is shared also with the $\varepsilon$ invariant tensors of the $\SU(N)$ groups, even though $d$ is symmetric, while the $\varepsilon$'s are antisymmetric under the exchanges of indices.

\section{The model}

\subsection{The general setup}

Our model is a renormalizable supersymmetric model, in which we spontaneously break $E_6$ to the SM gauge group. We shall not consider the orthogonal problem of SUSY breaking, so we will in fact end up with the MSSM.

Note first the decompositions of the $E_6$ representations into their $\SO(10)$ irreducible parts (with the $351'$ and $\overline{351'}$ exchanged compared to \cite{Slansky:1981yr}):
\begin{eqnarray}
27&=&16+10+1,\\
351'&=&1+10+16+54+126+\overline{144},
\end{eqnarray}
and analogously for their conjugate representations $\overline{27}$ and $\overline{351'}$.

We will use the fundamental representation $27$ to contain the Standard Model fermions: each generation will be present in the $16$ of $\SO(10)$, where the right-handed neutrino $\nu^c$ is also located. The remaining exotics are a vector-like pair of leptons and $d$-quarks in the $10$ (which is $5+\overline{5}$ under $\SU(5)$), and a Standard Model singlet $s$ in $1$ of $\SO(10)$, which is the analogue of a right-handed neutrino. We use the following intuitive notation for the vector-like exotics: $d'$, $d'^c$, $L'=(\nu',e')$ and $L'^c=(\nu'^c,e'^c)$. This simple picture is in reality complicated by mixing: the chiral SM fermions live in linear combination of primed and unprimed fields.

The model thus consists of $3$ copies of matter $27_F^i$ ($i=1,2,3$), and the Higgs sector, which consists of
\begin{eqnarray}
27+\overline{27}+351'+\overline{351'}+\widetilde{27}+\overline{\widetilde{27}}.
\end{eqnarray}
The non-tilde fields will acquire vacuum expectation values (VEV) at the GUT scale and will break $E_6$ directly to the Standard Model (no intermediate steps). It turns out that the tilde fields are needed in order to contain the MSSM Higgses, so they acquire VEVs at the electro-weak (EW) scale.

In contrast to the $\SO(10)$ group with large representations, where $R$-parity is automatically conserved, we need to impose in our model a global $\mathbf{Z}_2$ symmetry, under which
$27_F^i$ are odd. This enables us to prevent the fermionic $27$'s to acquire VEVs at the GUT scale.
Also, we take the tilde fields to couple to the non-tilde fields in the Higgs sector only in pairs, so that the tilde fields also do not acquire large VEVs, which simplifies the analysis of the equations of motion. Therefore, we have
\begin{eqnarray}
0&=&\langle 27_F^i\rangle_{{\rm GUT}}=\langle \widetilde{27}\rangle_{{\rm GUT}}=\langle \overline{\widetilde{27}}\rangle_{{\rm GUT}}.
\end{eqnarray}

\subsection{The Higgs sector}

\subsubsection{The $F$-terms}

The fields relevant for the breaking at the GUT scale are the non-tilde fields $27$, $\overline{27}$, $351'$ and $\overline{351'}$. These fields respectively contain $2$, $2$, $5$ and $5$ Standard Model singlets, which can acquire GUT-scale VEVs. The definition of our labels for these singlets can be found in Table~\ref{table:singlets}. The singlet VEVs have the standard K\" ahler normalization
\begin{eqnarray}
\langle 27^\mu\;27^*_\mu\rangle&=&|c_1|^2+|c_2|^2,\\
\langle\overline{27}_\mu\;\overline{27}^{*\mu} \rangle&=&|d_1|^2+|d_2|^2,\\
\langle 351'^{\mu\nu}\,351^{\prime *}_{\mu\nu} \rangle&=&|e_1|^2+|e_2|^2+|e_3|^2+|e_4|^2+|e_5|^2,\\
\langle\overline{351'}_{\mu\nu}\,\overline{351'}^{*\mu\nu} \rangle&=&|f_1|^2+|f_2|^2+|f_3|^2+|f_4|^2+|f_5|^2.
\end{eqnarray}
\begin{table}[h]
\begin{tabular}{crrr|crrr}
\hline
\tablehead{1}{r}{b}{VEV label}&
\tablehead{1}{r}{b}{$\subset\SU(5)$}&
\tablehead{1}{r}{b}{$\subset\SO(10)$}&
\tablehead{1}{r}{b}{$\subset \mathrm{E}_6$}&
\tablehead{1}{r}{b}{VEV label}&
\tablehead{1}{r}{b}{$\subset\SU(5)$}&
\tablehead{1}{r}{b}{$\subset\SO(10)$}&
\tablehead{1}{r}{b}{$\subset \mathrm{E}_6$}\\\hline
$c_1$&      $1$    &$1$                &$27$&$d_1$&$1$    &$1$                &$\overline{27}$\\
$c_2$&      $1$    &$16$               &$27$&$d_2$&$1$    &$\overline{16}$    &$\overline{27}$\\\hline
$e_1$&$1$&$126$&$351'$&$f_1$&$1$&$\overline{126}$&$\overline{351'}$\\
$e_2$&$1$&$16$&$351'$&$f_2$&$1$&$\overline{16}$&$\overline{351'}$\\
$e_3$&$1$&$1$&$351'$&$f_3$&$1$&$1$&$\overline{351'}$\\
$e_4$&$24$&$54$&$351'$&$f_4$&$24$&$54$&$\overline{351'}$\\
$e_5$&$24$&$\overline{144}$&$351'$&$f_5$&$24$&$144$&$\overline{351'}$\\\hline
\end{tabular}
\caption{The labels of Standard Model singlet VEVs in our model and their location in the embedding chain $\textrm{SM}\subset\SU(5)\subset\SO(10)\subset E_6$. \label{table:singlets}}
\end{table}

The most general renormalizable superpotential of the Higgs sector containing the non-tilde fields is
\begin{eqnarray}
W&=&m_{351'}\;I_{351'\times\overline{351'}}+m_{27}\;I_{27\times\overline{27}} + \lambda_1 \;I_{351'^3}+\lambda_2 \;I_{\overline{351'}^3}+\lambda_3 \;I_{27^2\times \overline{351'}}+\lambda_4 \;I_{\overline{27}^2\times 351'}+\lambda_5 \;I_{27^3} + \lambda_6 \;I_{\overline{27}^3}.\label{equation-superpotential}
\end{eqnarray}
The above invariants are explicitly computed to be
    \begin{eqnarray}
        I_{351'\times\overline{351'}}&=&\overline{351'}_{\mu\nu}\;351^{\mu\nu}=e_1 f_1+e_2 f_2+e_3 f_3+e_4 f_4+e_5 f_5,\\
        I_{27\times\overline{27}}&=&\overline{27}_\mu\; 27^\mu=c_1 d_1+c_2 d_2,\\
        I_{351'^3}&=&351'^{\mu\alpha}\;351'^{\nu\beta}\;351'^{\lambda\gamma}\;d_{\alpha\beta\gamma}d_{\mu\nu\lambda}=3 \left(e_3 e_4^2+e_1 e_5^2 - \sqrt{2} e_2 e_4 e_5 \right),\\
        I_{\overline{351'}^3}&=&\overline{351'}_{\mu\alpha}\;\overline{351'}_{\nu\beta}\;\overline{351'}_{\lambda\gamma}\;d^{\alpha\beta\gamma}\;d^{\mu\nu\lambda}=3 \left(f_3 f_4^2+f_1 f_5^2 - \sqrt{2}f_2 f_4 f_5\right),\\
        I_{27^2\times\overline{351'}}&=&\overline{351'}_{\mu\nu}\;27^\mu\;27^\nu=c_2^2 f_1 + \sqrt{2} c_1 c_2 f_2 + c_1^2 f_3,\\
        I_{\overline{27}^2\times 351'}&=&351'^{\mu\nu}\;\overline{27}_\mu\;\overline{27}_\nu=d_2^2 e_1 + \sqrt{2} d_1 d_2 e_2 + d_1^2 e_3,\\
        I_{27^3}&=&27^\mu\;27^\nu\;27^\lambda\;d_{\mu\nu\lambda}=0,\\
        I_{\overline{27}^3}&=&\overline{27}_\mu\;\overline{27}_\nu\;\overline{27}_\lambda\;d^{\mu\nu\lambda}=0.\label{equation-invariant-end}
    \end{eqnarray}
The zero of the invariants $27^3$ and $\overline{27}^3$ can be understood from the property of the $d$-tensor: the $27$ and $\overline{27}$ contain SM singlets only at two locations, so the singlet terms in the cubic invariant get $d$-tensor coefficients with at least two indices taking the same value, therefore these coefficients are zero.

Taking derivatives of the superpotential $W$ over all the different VEVs yields the $F$-terms.

\subsubsection{The $D$-terms}
In our model, the $D$-terms take the form

\begin{eqnarray}
D^A&=& (27^\dagger)_\mu\;(\hat{t}^A\, 27)^\mu+(\overline{27}^\dagger)^\mu\;(\hat{t}^A\, \overline{27})_\mu+(351'^\dagger)_{\mu\nu}\;(\hat{t}^A\, 351')^{\mu\nu}+
(\overline{351'}^\dagger)^{\mu\nu}\;(\hat{t}^A\, \overline{351'})_{\mu\nu},
\end{eqnarray}
where $\hat{t}^A$ is the action of the $A$-th generator of the $E_6$ algebra, and $A=1,\ldots,78$. Explicit computation shows that there are only $4$ independent non-zero real $D$-terms. They are

\begin{eqnarray}
D^{I}&=& |\RED{c_1}|^2 - |\RED{d_1}|^2 + |\RED{e_2}|^2 - |\RED{f_2}|^2 + 2 |\RED{e_3}|^2 - 2 |\RED{f_3}|^2 - |\RED{e_4}|^2 + |\RED{f_4}|^2,\\
D^{II}&=& |\RED{c_2}|^2 - |\RED{d_2}|^2 + |\RED{e_2}|^2 - |\RED{f_2}|^2 + 2 |\RED{e_1}|^2 - 2 |\RED{f_1}|^2 - |\RED{e_5}|^2 + |\RED{f_5}|^2,\\
D^{III}&=&
\RED{c_1} \RED{c_2}^\ast - \RED{d_1}^\ast \RED{d_2} + \sqrt{2} \RED{e_1}^\ast \RED{e_2} - \sqrt{2} \RED{f_1} \RED{f_2}^\ast + \sqrt{2} \RED{e_2}^\ast \RED{e_3} - \sqrt{2} \RED{f_2} \RED{f_3}^\ast + \RED{e_4}^\ast \RED{e_5} -
 \RED{f_4} \RED{f_5}^\ast,
\end{eqnarray}
where the last one is complex.
They respectively correspond to the following generators of the $\SU(3)_C\times\SU(3)_L\times\SU(3)_R$ subgroup of $E_6$: \hbox{$\sqrt{3}t_L^{8}+2t_R^{3}$,} \hbox{$-2 t_R^{3}$} and \hbox{$t_R^{6}+i\,t_R^{7}$}.

\subsubsection{Conjugation symmetry and the general solving strategy}

Our model is conjugate symmetric in the sense that our Higgs sector consists of pairs $R+\overline{R}$ of representations. The exchange of the representations with their conjugates, e.g.~$27\leftrightarrow\overline{27}$ and $351'\leftrightarrow\overline{351'}$, which can be more specifically written as $c_i\leftrightarrow d_i$ and $e_j\leftrightarrow f_j$, with $i=1,2$ and $j=1,\ldots,5$, will yield equivalent $D$-terms. But the $F$-terms change under this exchange, because the superpotential itself is not conjugation invariant. The reason for this are the coupling constants $\lambda_i$; only if we also exchange $\lambda_1\leftrightarrow \lambda_2$, $\lambda_3\leftrightarrow \lambda_4$ and $\lambda_5\leftrightarrow \lambda_6$ will the superpotential remain invariant.

The above fact that we have different parameters in front of invariants and their conjugate invariants, will have implications on our solving strategy for the equations of motion. One would perhaps be tempted to use the ansatz $\langle 27\rangle=\langle\overline{27}\rangle$ and $\langle 351'\rangle=\langle \overline{351'}\rangle$ to first get rid of the $D$-terms in a trivial manner, and then proceed to the $F$-terms. But the conjugate symmetric ansatz leads to
a consistent set of $F$-terms only if we assume an exact fine-tuning $\lambda_1=\lambda_2$ and $\lambda_3=\lambda_4$.

In the general case, assuming no relations among the superpotential parameters, it turns out the best strategy involves first solving the $F$-terms, and only then proceeding to solve the $D$-terms in a nontrivial manner.

\subsubsection{Solutions of the equations of motion}

There are many possible solutions to the equations of motion, the simplest of course being the trivial one with all VEVs zero.
Assuming $c_1,d_1\neq 0$ and $e_5,f_5\neq 0$, the $F$-terms are solved by the ansatz
\begin{eqnarray}
    d_1&=&\frac{m_{351'} m_{27}-2 \lambda_3 \lambda_4 \RED{c_2} \RED{d_2}}{2 \lambda_3 \lambda_4 \RED{c_1}},\label{general-first}\\
    e_1&=&-\frac{\lambda_3 \RED{c_2}^2+\frac{m_{351'}^2 (m_{351'} m_{27}-2 \lambda_3 \lambda_4 \RED{c_2} \RED{d_2})^2}{108 m_{27}^2 \lambda_1^2 \lambda_2 \RED{e_5}^2}}{m_{351'}},\\
    f_1&=&-\frac{\lambda_4 \RED{d_2}^2+3 \lambda_1 \RED{e_5}^2}{m_{351'}},\\
    e_2&=&\frac{\lambda_3 \RED{c_1} \left(m_{27} \lambda_4 \RED{d_2} m_{351'}^3-2 \lambda_3 \lambda_4^2 \RED{c_2} \RED{d_2}^2 m_{351'}^2-54 m_{27}^2 \lambda_1^2 \lambda_2 \RED{c_2} \RED{e_5}^2\right)}{27 \sqrt{2} m_{351'} m_{27}^2 \lambda_1^2 \lambda_2 \RED{e_5}^2},\\
    f_2&=&\frac{2 \lambda_3 \RED{c_2} \left(\lambda_4 \RED{d_2}^2+3 \lambda_1 \RED{e_5}^2\right)-m_{351'} m_{27} \RED{d_2}}{\sqrt{2} m_{351'} \lambda_3 \RED{c_1}},\\
    e_3&=&\frac{\lambda_3 \RED{c_1}^2 \left(-\frac{m_{351'}^2 \lambda_3 \lambda_4^2 \RED{d_2}^2}{m_{27}^2 \lambda_1^2 \lambda_2 \RED{e_5}^2}-27\right)}{27 m_{351'}},\\
    f_3&=&-\frac{m_{351'}^2 m_{27}^2-4 m_{351'} \lambda_3 \lambda_4 \RED{c_2} \RED{d_2} m_{27}+4 \lambda_3^2 \lambda_4 \RED{c_2}^2 \left(\lambda_4 \RED{d_2}^2+3 \lambda_1 \RED{e_5}^2\right)}{4 m_{351'} \lambda_3^2 \lambda_4 \RED{c_1}^2},\\
    e_4&=&\frac{\RED{c_2} \RED{e_5}}{\RED{c_1}},\\
    f_4&=&\frac{m_{351'} \lambda_3 \lambda_4 \RED{c_1} \RED{d_2}}{9 m_{27} \lambda_1 \lambda_2 \RED{e_5}},\\
    f_5&=&\frac{m_{351'} (m_{351'} m_{27}-2 \lambda_3 \lambda_4 \RED{c_2} \RED{d_2})}{18 m_{27} \lambda_1 \lambda_2 \RED{e_5}}.\label{general-last}
\end{eqnarray}
The remaining VEVs $c_1$, $e_5$, $c_2$ and $d_2$ are then determined by the $4$ $D$-terms. One possible solution is to take
\begin{eqnarray}
\label{c2d2zero}
c_2&=&d_2=0,\\
\label{e5}
e_5&=& \frac{m_{351'}}{3 \sqrt{2} \lambda_1^{2/3} \lambda_2^{1/3}},
\end{eqnarray}
which then gives the following polynomial condition for $c_1$:
\begin{eqnarray}
    0&=& |m_{351'}|^4 |m_{27}|^4 + 2 |m_{351'}|^4 |m_{27}|^2 |\lambda_3|^2 |\RED{c_1}|^2-8 |m_{351'}|^2 |\lambda_3|^4 |\lambda_4|^2 |\RED{c_1}|^6 - 16 |\lambda_3|^6 |\lambda_4|^2 |\RED{c_1}|^8.
\end{eqnarray}
Note that the form of the polynomial ensures that there always exists a solution $c_1>0$.

This solution breaks $E_6$ to the Standard Model, which we determined by the computation of gauge boson masses, $12$ of which remain zero. The only alternative ansatz for the $F$-terms, which also breaks to the Standard Model, is an analogue of equations~\eqref{general-first}-\eqref{general-last}, where we exchange
$c_1\leftrightarrow d_1$, $c_2\leftrightarrow d_2$, $e_1\leftrightarrow e_3$, $f_1\leftrightarrow f_3$, $e_4\leftrightarrow e_5$ and $f_4\leftrightarrow f_5$. This ansatz assumes $c_2,d_2\neq 0$ and $e_4,f_4\neq 0$.

All other solutions of the equations of motion do not break to the Standard Model group. In fact, all but one of the other solutions leave $\SU(5)$ unbroken. The exception involves taking the ansazt $\langle 27\rangle=\langle \overline{27}\rangle=0$ and solving the $F$-terms; by computing which gauge bosons remain massless, we find that this scenario breaks to a Pati-Salam group, which is embedded into $E_6$ in a non-canonical way. Note that the scenario $\langle 27\rangle=\langle \overline{27}\rangle=0$ corresponds to having a model with the Higgs sector $351'+\overline{351'}$. This shows why we add the extra $27+\overline{27}$ pair to the breaking sector.

\subsection{The Yukawa sector}

The Yukawa sector in our model is composed from the following terms:
\begin{eqnarray}
\mathcal{L}_{{\rm Yukawa-}E_6}&=&\frac{1}{2}\;27_F^i\; 27_F^j\;(Y_{27}^{ij}\; 27 + Y_{\overline{351'}}^{ij} \;\overline{351'}+\widetilde{Y}_{27}^{ij} \;\widetilde{27})\label{equation:Yukawa}.
\end{eqnarray}
The model contains three Yukawa mixing matrices: $Y_{27}$, $Y_{\overline{351'}}$ and $\widetilde{Y}_{27}$.

Compare this with the Yukawa terms in the renormalizable SUSY $\SO(10)$ model:
\begin{eqnarray}
\mathcal{L}_{\rm Yukawa-\SO(10)}&=&\frac{1}{2}16_F^i\; 16_F^j\;(Y_{10}^{ij}\; 10 + Y_{\overline{126}}^{ij} \;\overline{126}).
\end{eqnarray}
The analogy between the two models is not accidental. The fermionic $27_F^i$ in $E_6$ are analogous to the fermionic $16_F^i$ in $\SO(10)$, while $27$ and $\overline{351'}$ of $E_6$ function as analogues of $10$ and $\overline{126}$, respectively. This is not surprising, since the $27$ contains both the $16$ and $10$ of $\SO(10)$, while $\overline{351'}$ contains the $\overline{126}$ of $\SO(10)$, so the Yukawa terms of the $\SO(10)$ model will be automatically included also in our model. But the $E_6$ Yukawa terms contain also terms involving the exotics, such as \hbox{$16_F^i 10_F^j (Y_{27}^{ij}\; 16+Y_{\overline{351'}}^{ij}\; 144)$} in $\SO(10)$ language.

The mechanism of flavor mixing differs in the two models significantly.

In the $\SO(10)$ model, we need two (symmetric) Yukawa matrices for flavor mixing, since a single one can always be diagonalized with a redefinition of generations. The SM Higgs boson is located in both the $10$ and the $\overline{126}$, its EW scale mass ensured by a fine tuning of parameters, such that one weak Higgs doublet mode is (almost) massless, while all color triplets remain heavy.

In the $E_6$ model, the fields $27$ and $\overline{351'}$ acquire first GUT scale VEVs, which causes vector-like pairs of quarks and leptons to acquire heavy masses. In $\SU(5)$ language, the heavy $\overline{5}$ is a linear combination of the $\overline{5}$'s in the $16$ and $10$ contained in $27_F^i$. One would then hope to make a combination of the doublets in the representations $27$, $\overline{351'}$ and possibly $351'$ and $\overline{27}$ massless, while keeping the color triplets heavy.

There are $3$ doublet/antidoublet pairs and $3$ triplet/antitriplet pairs in the $27+\overline{27}$ pair, while $351'+\overline{351'}$ have $8$ doublet/antidoublet pairs and $9$ triplet/antitriplet pairs. The doublet and triplet mass matrices therefore have dimensions $11\times 11$ and $12\times 12$, respectively. Both matrices automatically have a massless mode, which correspond to the would-be Nambu-Goldstone bosons of the $E_6$ breaking. Doublet-triplet splitting in this case therefore means a fine-tuning of parameters, such that the doublets get an extra massless mode, while the triplets don't. The explicit conditions for the extra massless mode of doublets and triplets, after using any of the solutions of the equations of motion, which break to the Standard Model, give
\begin{eqnarray}
\mathrm{Cond}_{{\rm doublets}}&=&\frac{m_{351'}^9 m_{27} \lambda_3 \lambda_4}{72 \lambda_1 \lambda_2},\\
\mathrm{Cond}_{{\rm triplets}}&=&\frac{4 m_{351'}^{10} m_{27} \lambda_3 \lambda_4}{243 \lambda_1 \lambda_2}.
\end{eqnarray}
It is not possible to fine-tune the parameters ($m$'s and $\lambda$'s) so that a doublet mode gets massless, but a triplet doesn't.

The inability to perform doublet-triplet splitting in the non-tilde sector is the reason for adding the tilde fields $\widetilde{27}+\overline{\widetilde{27}}$. We assume that they combine with the non-tilde fields only in pairs, so they needn't acquire GUT scale VEVs, but only EW VEVs. We now have new parameters from terms with fields, which are not involved in the symmetry breaking, and we denote them by $\kappa$. The terms with the tilde fields in the superpotential are
\begin{eqnarray}
m_{\widetilde{27}}\;\widetilde{27}\;\overline{\widetilde{27}}+\kappa_1\;\widetilde{27}\;\widetilde{27}\;\overline{351'}+\kappa_2\;\overline{\widetilde{27}}\;\overline{\widetilde{27}}\;351'+\kappa_3 \; \widetilde{27}\;\widetilde{27}\;27+\kappa_4 \; \overline{\widetilde{27}}\;\overline{\widetilde{27}}\;\overline{{27}}.
\end{eqnarray}
The above terms enable fine-tuning in the tilde sector, so that doublets in the $\widetilde{27}$ and $\overline{\widetilde{27}}$ acquire an EW VEV.
We label the VEVs in the $\widetilde{27}$ by $u_1$, $v_1$ and $v_2$ (more details can be found in Table~\ref{table:doublets}). Notice that
$u_1$ and $(|v_1|^2+|v_2|^2)^{1/2}$ are not the MSSM Higgs VEVs yet: parts of them lie also in $\overline{\widetilde{27}}$. In the following, we will assume that these VEVs can all be non-zero; this indeed turns out to be the case, which is shown in the section on doublet-triplet splitting.

An explicit calculation of the the Yukawa terms in equation~\eqref{equation:Yukawa} then gives
\begin{eqnarray}
u^{T}(-u_1)\ydt u^c +
\pmatrix{
d^{cT} & d'^{cT}\cr
}
\pmatrix{
\phantom{-}v_1\ydt & \phantom{-}c_2\yd+\frac{f_5}{\sqrt{15}}\yt \cr
-v_2\ydt & -c_1\yd+\frac{f_4}{\sqrt{15}}\yt \cr
}\!\!
\pmatrix{
d \cr
d'
}+
\pmatrix{
e^{T} & e'^{T}
}
\pmatrix{
-v_1\ydt & \phantom{-}c_2\yd-\frac{3}{2}\frac{f_5}{\sqrt{15}}\yt \cr
\phantom{-}v_2\ydt & -c_1\yd-\frac{3}{2}\frac{f_4}{\sqrt{15}}\yt
}\!\!
\pmatrix{
e^c \cr
e'^c
}&&\nonumber\\
+
\pmatrix{
\nu^{T} & \nu'^{T}
}\!
\pmatrix{
u_1\ydt & 0 & \phantom{-}c_2\yd-\frac{3}{2}\frac{f_5}{\sqrt{15}}\yt \cr
0 & -u_1\ydt & -c_1\yd-\frac{3}{2}\frac{f_4}{\sqrt{15}}\yt
}\!\!
\pmatrix{
\nu^c \cr
s \cr
\nu'^c
}
+
\pmatrix{
\nu^{cT} & s^T & \nu'^{cT}
}\!
\pmatrix{
f_1\yt & \frac{f_2}{\sqrt{2}}\yt & -v_2\ydt \cr
\frac{f_2}{\sqrt{2}}\yt & f_3\yt & \phantom{-}v_1\ydt \cr
-v_2\ydt & v_1\ydt & 0
}\!\!
\pmatrix{
\nu^c \cr
s \cr
\nu'^c
}\!.
\end{eqnarray}
Using the standard techniques for integrating out the heavy vector-like families (see for example
\cite{Babu:2012pb} and references therein) one arrives in the case (\ref{c2d2zero}), (\ref{e5}),
at the following expressions for the Yukawa matrices:
\bea
M_D^T&=&\left(1+(4/9)XX^\dagger\right)^{-1/2}\left(v_1-(2/3)v_2X\right)\ydt,\\
M_E&=&-\left(1+XX^\dagger\right)^{-1/2}\left(v_1+v_2X\right)\ydt,\\
M_U&=&-u_1\ydt,\\
M_N&=&\left(1+XX^\dagger\right)^{-1/2}\left(\frac{u_1^2}{f_1}\ydt\yt^{-1}\ydt+
\frac{u_1^2}{f_3}X\ydt\yt^{-1}\ydt X^T\right)\left(1+X^*X^T\right)^{-1/2},
\eea
where the matrix $X$ is defined as
\beq
X\equiv \sqrt{\frac{3}{20}}\frac{f_5}{c_1}\yt\yd^{-1}.
\eeq
and where only the type I seesaw contribution has been taken into account for simplicity.

The number of parameters involved seems to easily accommodate the light fermion masses.
In fact two symmetric matrices are typically able to describe the charged fermion sector (see for example
the SO(10) case with 10 and 126 Higgses), while a third matrix should easily take care of the neutrino
sector.

\subsection{Details on doublet-triplet splitting}

Doublet triplet splitting is possible only in the tilde sector. The relevant doublet/antidoublet and triplet/antitriplet mass matrices in this sector are
\begin{eqnarray}
\widetilde{M}_{\textrm{doublets}}&=&\pmatrix{
 m_{\widetilde{27}} & -2 \kappa_3 \RED{c_1}-3\kappa_1 \frac{\RED{f_4}}{\sqrt{15}} & 2 \kappa_3\RED{c_2}-3 \kappa_1 \frac{\RED{f_5}}{\sqrt{15}} \cr
 -2 \kappa_4 \RED{d_1}-3 \kappa_2 \frac{\RED{e_4}}{\sqrt{15}} & m_{\widetilde{27}} & 0 \cr
 \phantom{-}2 \kappa_4 \RED{d_2}-3  \kappa_2 \frac{\RED{e_5}}{\sqrt{15}} & 0 & m_{\widetilde{27}}
},\\
\widetilde{M}_{\textrm{triplets}}&=&\pmatrix{
m_{\widetilde{27}} & -2 \kappa_3\RED{c_1}+2\kappa_1 \frac{\RED{f_4}}{\sqrt{15}} & 2 \kappa_3 \RED{c_2}+2 \kappa_1 \frac{\RED{f_5}}{\sqrt{15}} \cr
 -2 \kappa_4 \RED{d_1}+2 \kappa_2 \frac{\RED{e_4}}{\sqrt{15}} & m_{\widetilde{27}} & 0 \cr
 \phantom{-}2 \kappa_4 \RED{d_2}+2 \kappa_2 \frac{\RED{e_5}}{\sqrt{15}} & 0 & m_{\widetilde{27}}
},
\end{eqnarray}
\noindent
with the mass terms being
\begin{eqnarray}
\pmatrix{D_1^T&D_2^T&D_3^T}\;\widetilde{M}_{{\rm doublets}}\;\pmatrix{\overline{D}_1\cr \overline{D}_2\cr \overline{D}_3}+\quad\pmatrix{T_1^T&T_2^T&T_3^T}\;\widetilde{M}_{{\rm triplets}}\; \pmatrix{\overline{T}_1\cr \overline{T}_2\cr \overline{T}_3}.
\end{eqnarray}
The labels $D$, $T$ generically denote doublets $(1,2,\frac{1}{2})$ and triplets $(3,1,-\frac{1}{3})$ in $5$'s of $\SU(5)$, while $\overline{D}$ and $\overline{T}$ denote the antidoublets $(1,2,-\frac{1}{2})$ and antitriplets $(\overline{3},1,\frac{1}{3})$ in $\overline{5}$'s of $\SU(5)$. More details on the doublets and triplets are shown in Table~\ref{table:doublets}. Note that the EW VEVs $u_1$, $v_1$ and $v_2$ correspond to the fields $D_1$, $\overline{D}_2$ and $\overline{D}_3$, respectively. While $D_2$, $D_3$ and $\overline{D}_1$ also acquire VEVs, we will not label them.
\begin{center}
\begin{table}[ht]
\begin{tabular}{lllll}
\hline
doublet,triplet&$\subset\SU(5)$&$\subset\SO(10)$&$\subset E_6$&doublet VEV\\\hline
$D_1,T_1$&$5$&$10$&$\widetilde{27}$&$u_1$\\
$D_2,T_2$&$5$&$10$&$\overline{\widetilde{27}}$&\\
$D_3,T_3$&$5$&$\overline{16}$&$\overline{\widetilde{27}}$&\\\hline
$\overline{D}_1,\overline{T}_1$&$\overline{5}$&$10$&$\overline{\widetilde{27}}$&\\
$\overline{D}_2,\overline{T}_2$&$\overline{5}$&$10$&$\widetilde{27}$&$v_1$\\
$\overline{D}_3,\overline{T}_3$&$\overline{5}$&$16$&$\widetilde{27}$&$v_2$\\\hline
\end{tabular}
\caption{Labels of the doublets and triplets along with their locations in $\widetilde{27}$ and $\overline{\widetilde{27}}$.\label{table:doublets}}
\end{table}
\end{center}

Plugging the vacuum solution into the mass matrices for doublets and triplets, we get the following conditions for doublet triplet spliting:
\begin{eqnarray}
0&=& m_{\widetilde{27}}^3 - \frac{1}{30}\;\;m_{\widetilde{27}} m_{351'}^2 \frac{\kappa_1 \kappa_2}{\lambda_1 \lambda_2} - 2 m_{\widetilde{27}} m_{351'} m_{27} \frac{\kappa_3 \kappa_4}{\lambda_3 \lambda_4},\\
0&\neq& m_{\widetilde{27}}^3 - \frac{2}{135}\; m_{\widetilde{27}} m_{351'}^2 \frac{\kappa_1 \kappa_2}{\lambda_1 \lambda_2} - 2 m_{\widetilde{27}} m_{351'} m_{27} \frac{\kappa_3 \kappa_4}{\lambda_3 \lambda_4}.
\end{eqnarray}
Both conditions are satisfied by a fine-tuning
\begin{eqnarray}
\kappa_1 &\approx& 30 (m_{\widetilde{27}}^2 \lambda_3 \lambda_4 - 2 m_{351'} m_{27} \kappa_3 \kappa_4)\frac{\lambda_1 \lambda_2 }{m_{351'}^2 \lambda_3 \lambda_4 \kappa_2}.
\end{eqnarray}
This fine-tuning gives the following modes of doublets and antidoublets to be massless:
\begin{eqnarray}
D_{m=0}&\propto&\frac{\sqrt{1/30}\;m_{\widetilde{27}} m_{351'} \lambda_1^{-2/3} \lambda_2^{-1/3} \lambda_3 \lambda_4 \kappa_2}{
 m_{\widetilde{27}}^2 \lambda_3 \lambda_4 -
    2 m_{351'} m_{27} \kappa_3 \kappa_4}\;D_1 + \frac{\sqrt{2/15}\;m_{351'}\RED{c_1} \lambda_1^{-2/3} \lambda_2^{-1/3}\lambda_3 \lambda_4 \kappa_2 \kappa_3}{m_{\widetilde{27}}^2 \lambda_3 \lambda_4 - 2 m_{351'} m_{27} \kappa_3 \kappa_4} \;D_2 + D_3,\\
\overline{D}_{m=0}&\propto&\frac{\sqrt{30}\;m_{\widetilde{27}} \lambda_1^{2/3} \lambda_2^{1/3}}{m_{351'} \kappa_2}\;\overline{D}_1+ \frac{\sqrt{30}\; m_{27} \lambda_1^{2/3} \lambda_2^{1/3} \kappa_4}{\RED{c_1} \lambda_3 \lambda_4 \kappa_2}\;\overline{D}_2+\overline{D}_3.
\end{eqnarray}
Notice that he massless modes have components of all doublets and antidoublets (in particular, $D_1$, $\overline{D}_2$ and $\overline{D}_3$ in $\widetilde{27}$). Since only the massless modes can acquire a large EW VEV (after SUSY breaking), the above fact ensures non-zero $u_1$, $v_1$ and $v_2$, which was assumed in the analysis of the Yukawa sector.

\section{Conclusion}

There is always a clash in physical models between being realistic and being predictive. Most of the times
theories are either realistic but knot predictive or predictive but wrong. In $E_6$ grand unified theories we are not
yet at a stage of being both realistic and predictive. We presented here an example of a realistic
$E_6$ theory: a renormalizable supersymmetric case with $3\times 27_F+2\times(27+\overline{27})+
351'+\overline{351}'$. This is important as an existence proof, the next step would be to try to find out
more minimal models, especially with regard to the number of Yukawa matrices. Our goal is to bring the $E_6$
GUTs at the level of the well studied SU(5) and SO(10) cases.




\begin{theacknowledgments}
We would like to thank Kaladi Babu, Ilia Gogoladze, St\'ephane Lavignac and
Zurab Tavartkiladze for discussion and correspondence. This work has been supported in part
by the Slovenian Research Agency. BB would like to thank the organizers of the
theory workshop on "Neutrino Physics and Astrophysics", July 15-26, 2013, Lead, South Dakota and
CETUP* (Center for Theoretical Underground Physics and Related Areas),
supported by the US Department of Energy under Grant No. DE-SC0010137 and by the US National Science
Foundation under Grant No. PHY-1342611, for its hospitality and partial support during the 2013 Summer Program.
\end{theacknowledgments}



\bibliographystyle{aipproc}   


\end{document}